\documentstyle[12pt]{article}
\begin{document}
\begin{center}
\bf{CURVATURE INHERITANCE SYMMETRY IN RIEMANNIAN SPACES WITH
APPLICATIONS TO STRING CLOUD AND STRING FLUIDS}\\[1cm]
\.{I}. Y\i lmaz$^{1}$, \.{I}. Yavuz$^{2}$, H. Baysal$^{3}$, \.{I}. Tarhan$
^{1}$ and U. Camc{\i }$^{3}$ \\[1cm]
$^{1}$Department of Physics, Faculty of Arts and Sciences, \c{C}anakkale
Onsekiz Mart University, 17100 \c{C}anakkale, Turkey.\\
e-mail: iyilmaz@hotmail.com and tarhan\_ism@hotmail.com\\[0.5cm]
$^{2}$Department of Computer Engineering, Faculty of Engineering and
Architecture, \c{C}anakkale Onsekiz Mart University, 17100 \c{C}anakkale,
e-mail: yavuz@hotmail.com\\[0.5cm]
$^{3}$Department of Mathematics, Faculty of Arts and Sciences, \c{C}anakkale
Onsekiz Mart University, 17100 \c{C}anakkale, Turkey.\\
e-mail: hbaysal@hotmail.com and ucamci@hotmail.com\\[2cm]
\end{center}
\begin{abstract}
We study, in this paper, curvature inheritance symmetry (CI), $\pounds
_{\xi}R_{bcd}^{a}=2\alpha R_{bcd}^{a}$, where $\alpha $ is a scalar
function, for string cloud and string fluid in the context of general
relativity. Also, we have obtained some result when a proper CI (i.e.,
$\alpha \neq 0$) is also a conformal Killing vector.\newline\\[1cm]
\end{abstract}

\pagestyle{plain}
\newpage
\section{Introduction}

The question of symmetry inheritance is concerned with determining when the
symmetries of geometry ( defined through the existence of symmetry vectors)
are inherited by the source terms or individual physical components of the
energy-stress tensor (related to the geometry via Einstein field equation).

The most useful inheritance symmetry is the symmetry under the conformal
motions (Conf M). A $V_n$ admits a Conf M generated by a conformal Killing
vector (CKV) $\xi $ if
\begin{equation}
\pounds _\xi g_{ab}=2\psi g_{ab},\,\psi
=\psi (x^a).
\end{equation}
It follows from Yano$^1$ that every Conf M must satisfy

\begin{equation}
\pounds _\xi \Gamma _{bc}^a=\delta _b^a\psi _{;c}+\delta _c^a\psi
_{;b}-g_{bc}g^{ad}\psi _{;d}
\end{equation}
\begin{equation}
\pounds _\xi R_{ab}=-g_{ab}\Box \psi -(n-2)\psi _{;ab}
\end{equation}
\begin{equation}
\pounds _\xi R=-2(n-1)\Box \psi -2\psi R
\end{equation}
\begin{equation}
\pounds _\xi C_{bcd}^a=0
\end{equation}
where $\pounds _\xi $ signifies the Lie derivative along $\xi ^a$ and $\psi
(x^a)$ is the conformal factor and $\Box $ is the Laplacian operator defined
by $\Box \psi =g^{ab}\psi _{;ab}$. In particular, $\xi $ is special
conformal Killing vector (SCKV) if $\psi _{;ab}=0$ and $\psi _{,a}\neq 0$.
Other subcases are homothetic vector (HV) if $\psi _{,a}=0$ and Killing
vector (KV) if $\psi =0$. Here (;) and (,) denote the covariant and ordinary
derivatives, respectively.

The study of inheritance symmetries with CKV's and SCKV in fluid
space-times (perfect, anisotropic, viscous and heat-conducting) has recently
attracted some interest. Herrera et al.$^2$ have studied CKV's, with
particular reference to perfect and anisotropic fluids; Mason and
Tsamparlis$^3$ have investigated spacelike CKV's; Maartens et al.$^4$
have made a study of CKV's in anisotropic fluids, in which they are
particularly concerned with special conformal Killing vector (SCKV); Coley
and Tupper$^5$ have discussed space-times admitting SCKV and symmetry
inheritance. Yavuz and Y{
and special conformal Killing vectors in string cosmology. Carot et
al.$^7$ have discussed space-times with conformal Killing vectors. Also,
Duggal$^{8,9}$ have discussed curvature inheritance symmetry in Riemannian
spaces with applications to fluid space-times. Recently, the Einstein
field equations for inhomogeneous cylindrically symmetric space-times
filled cosmic strings have been discussed by K{\i }l{\i }n\d c and
Yavuz$^{10}$. Some string dust models for Bianchi Type I space-time have
been studied in detail by Yavuz and Tarhan$^{11}$. The Einstein's field
equations for a cloud of string with heat flux in Bianchi Type III
space-time solved by Yavuz and Y{\i }lmaz $^{12}$. Bianchi Type I universes
representing different forms of material distributions have been studied
by Bali and Jain$^{13}$, Bali and Tyagi $^{14}$, Bali $^{15}$, Singh and
Srivastava $^{16}$, Bali and Deepak$^{17}$, K{\i }l{\i }n\d c$^{18}$
(see also references therein).

In this paper, we will examine curvature inheritance symmetry in the
space-times with string source (string cloud and string fluid).
Energy -momentum tensor for a cloud of strings can be written as

\begin{equation}
T_{ab}=\rho u_au_b-\lambda x_ax_b
\end{equation}
where $\rho $ is the rest energy for cloud of strings with particles
attached to them and $\lambda $ is string tensor density and are related by

\begin{equation}
\rho =\rho _p+\lambda .
\end{equation}
Here $\rho _p$ is particle energy density. The unit timelike vector $u^a$
describes the cloud four-velocity and the unit spacelike vector $x^a$
represents a direction of anisotropy, i.e., the string's directions$^{19}$.
We have

\begin{equation}
u^au_a=-x^ax_a=-1\,\,\,\,and\,\,\,\,u^ax_a=0.
\end{equation}
The energy-momentum tensor for a fluid of strings$^{20,21}$ is

\begin{equation}
T_{ab}=(q+\rho _s)(u_au_b-x_ax_b)+qg_{ab}
\end{equation}
Also, note that

\begin{equation}
u^au_a=-x^ax_a=-1\,\,\,\,and\,\,\,\,u^ax_a=0
\end{equation}
where $\rho _s$ is string density and $q$ is ''string tension'' and also
''pressure''.

The paper may be outlined as follows. In Section II curvature inheritance
equations in the cloud of string and in Section III equations state of
string cloud are obtained. In Section IV curvature inheritance equations in
the fluids of string are derived. In Section V the results are discussed.

\section{Curvature Inheritance Symmetry
In The Cloud of Strings}

Consider a Riemannian space $V_n$ of arbitrary signature. We define symmetry
called ''curvature inheritance'' (CI) on $V_n$ by an infinitesimal
transformation $\overline{x}^{a}=x^a+\xi ^a(x)\delta (t)$, for which

\begin{equation}
\pounds _\xi R^{a}_{bcd}=2\alpha R^{a}_{bcd}
\end{equation}
where $\alpha=\alpha(x)$ is a scalar function, $\delta(t)$ is a positive
infinitesimal and $R^{a}_{bcd}$ is the Riemannian curvature tensor defined by

\begin{equation}
R_{bcd}^{a}=\Gamma _{bd,c}^{a}-\Gamma _{bc,d}^{a}+\Gamma _{bd}^{e}\Gamma
_{ec}^{a}-\Gamma _{bc}^{e}\Gamma _{ed}^{a}
\end{equation}
Here $\Gamma _{bc}^{a}$ are the Christoffel symbols of the second kind. A
subcase of CI is the well-known symmetry ''curvature collineation'' (CC)
when $\alpha =0$. In the sequel, we say that CI is proper if $\alpha \neq 0$%
. If a $V_{n}$ admits a CI, then the following identities hold ($\pounds
_{\xi }g_{ab}\equiv h_{ab}$):

\begin{equation}
\pounds _{\xi }R_{ab}=2\alpha R_{ab},
\end{equation}
i.e., $\xi $ defines Ricci inheritance symmetry,

\begin{equation}
\pounds _{\xi }R_{b}^{a}=2\alpha R_{b}^{a}-R_{b}^{c}h_{c}^{a},
\end{equation}
\begin{equation}
\pounds _{\xi }R=2\alpha R-R^{\prime },
\end{equation}
\begin{equation}
\pounds _{\xi }W_{bcd}^{a}=2\alpha W_{bcd}^{a},
\end{equation}
\begin{equation}
\pounds _{\xi }C_{bcd}^{a}=2\alpha C_{bcd}^{a}+D_{bcd}^{a},
\end{equation}
where $R$ is the scalar curvature, $R^{\prime }=R_{ab}h^{ab}$ and %
\begin{equation}
W_{bcd}^{a}=R_{bcd}^{a}-\frac{1}{n-1}\left[ \delta _{d}^{a}R_{bc}-\delta
_{c}^{a}R_{bd}\right] \,\,{\small :Weyl\,\,projective\,\,tensor,}
\end{equation}

\begin{eqnarray}
C_{bcd}^{a}&=&R_{bcd}^a+\frac {1}{n-2}\left[ \delta _c^aR_{bd}-\delta
_d^aR_{bc}+g_{bd}R_c^a-g_{bc}R_d^a\right] \nonumber \\
& &+\frac R{{(n-1)(n-2)}}\left[ \delta _d^ag_{bc}-\delta _c^ag_{bd}\right]\,
\small{:Conformal\,\,curvature\,\,tensor,}
\end{eqnarray}

\begin{eqnarray}
D_{bcd}^a&=&\frac 1{n-2}\left[
h_{bd}R_c^a-h_{bc}R_d^a+g_{bc}R_d^eh_e^a-g_{bd}R_c^eh_e^a\right] \nonumber \\
& &+\frac 1{{(n-1)(n-2)}}\left[ \delta _d^a(Rh_{bc}-R^{\prime }g_{bc})-\delta
_c^a(Rh_{bd}-R^{\prime }g_{bd})\right]
\end{eqnarray}
A CI is also a CKV$^{7,8}$ if

\begin{equation}
\psi _{;ab}=\frac{\alpha }{n-2}\left[ \frac{R}{n-1}g_{ab}-2R_{ab}\right]
\end{equation}
\begin{equation}
\Box \psi +\frac{\alpha R}{n-1}=0
\end{equation}
\begin{equation}
\alpha =\psi +\xi ^{a}\partial _{a}(log\sqrt{R})
\end{equation}

The proof of Eq. (21) follows by comparing (13), (15) with (3).
Also, the proof of Eq. (22) follows from (4) and (21). If $X^a$ is any
unit vector (timelike or spacelike) and $\xi^{a}$ is a CKV satisfying (1),
then

\begin{equation}
\pounds _{\xi }X^{a}=-\psi X^{a}+Y^{a},
\end{equation}
\begin{equation}
\pounds _{\xi }X_{a}=\psi X_{a}+Y_{a},
\end{equation}
where $Y^{a}$ is some vector orthogonal to $X^{a}$, i.e., $X^{a}Y_{a}=0$,
(see Ref. 3).

Applying the results (24) and (25) to the timelike unit four-velocity
vector $u^a$ and to spacelike unit vector $x^a$ of the string cloud and
string fluid, we have

\begin{equation}
\pounds _\xi u^a=-\psi u^a+v^a,
\end{equation}

\begin{equation}
\pounds _\xi u_a = \psi u_a + v_a ,
\end{equation}
where $u_a v^a = 0$ and

\begin{equation}
\pounds _\xi x^a = -\psi x^a + n^a ,
\end{equation}

\begin{equation}
\pounds _\xi x_a = \psi x_a + n_a ,
\end{equation}
where $x_a n^a = 0$. Since $x_a u^a = 0$ [see eq.(8)] we have

\begin{equation}
x_a\pounds _\xi u^a+u^a\pounds _\xi x_a=0.
\end{equation}
Substituting Eqs. (26) and (29) into (30), we get

\begin{equation}
v_ax^a+n_au^a=0.
\end{equation}
If $\xi ^a$ is a CKV satisfying (1), then

\begin{equation}
\pounds _{_\xi }R_{ab}=-2\psi _{;ab}-g_{ab}\Box \psi ,
\end{equation}

\begin{equation}
\pounds _{_\xi }R=-2\psi R-6\Box \psi ,
\end{equation}

\begin{equation}
\pounds _{_\xi }G_{ab}=2g_{ab}\Box \psi -2\psi _{;ab},
\end{equation}
where $\Box \psi \equiv \ g^{ab}\psi _{;ab},\,\,R_{ab}$ is Ricci tensor, $%
G_{ab}$ is Einstein tensor and $R=g^{ab}R_{ab}$ is Ricci scalar. Via
Einstein's field equations
\begin{equation}
G_{ab}\equiv R_{ab}-\frac 12Rg_{ab}=T_{ab}
\end{equation}
we find for $T_{ab}$
\begin{equation}
\pounds _{_\xi }T_{ab}=2g_{ab}\Box \psi -2\psi _{;ab}
\end{equation}
We take $T_{ab}$ to be of the form (6). With the aid of (27) for $%
\pounds _{_\xi }u_a$ and (29) for $\pounds _{_\xi }x_a$ a direct
calculation yields

\begin{equation}
\pounds _{_\xi }T_{ab}=[\pounds _\xi \rho +2\psi \rho ]u_au_b-[\pounds
_{_\xi }\lambda +2\psi \lambda ]x_ax_b+2\rho u_{(a}v_{b)}-2\lambda
x_{(a}n_{b)},
\end{equation}
which, when substituted into (36), gives
\begin{eqnarray}
2\Box \psi (h_{ab}-u_au_b+x_ax_b)-2\psi _{;ab}&=&[\pounds _{_\xi }\rho +2\psi
\rho ]u_au_b-[\pounds _{_\xi }\lambda +2\psi \lambda ]x_ax_b \nonumber \\
& &+2\rho u_{(a}v_{b)}-2\lambda x_{(a}n_{b)}
\end{eqnarray}
where $h_{ab}\,$ is projection tensor that projects in the directions that
are perpendicular to both $x^a$ and $u^a$,

\begin{equation}
h_{ab}=g_{ab}+u_au_b-x_ax_b.
\end{equation}
Some properties of this tensor are

\begin{equation}
h^{ab}u_{b}=h^{ab}x_{b}=0
\end{equation}

\begin{equation}
h_{c}^{a}h_{b}^{c}=h_{b}^{a},\,\,h_{ab}=h_{ba}
\end{equation}
By contracting (38) with the tensors $%
u^{a}u^{b},x^{a}x^{b},u^{a}x^{b},u^{a}h^{bc}$ and $x^{a}h^{bc}$ the
following equations for the string cloud are derived:

\begin{equation}
\pounds _{_{\xi }}\rho +2\psi \rho =-2(\Box \psi +\psi _{;ab}u^{a}u^{b})
\end{equation}

\begin{equation}
\pounds _{_\xi }\lambda +2\psi \lambda =2(\Box \psi -\psi _{;ab}x^ax^b)
\end{equation}

\begin{equation}
(\rho -\lambda )x^bv_b=\rho _px^bv_b=2\psi _{;ab}u^ax^b
\end{equation}

\begin{equation}
\rho h^{bc}v_b=2\psi _{;ab}u^ah^{bc}
\end{equation}

\begin{equation}
\lambda h^{bc}n_b=2\psi _{;ab}x^ah^{bc}
\end{equation}
Equations (42)-(46) are valid for any CKV $\xi ^a$. For the string
cloud, using the Einstein equations (35), we obtain

\begin{equation}
R_{ab}u^{a}u^{b}=\rho -R/2,\,\,R_{ab}x^{a}x^{b}=R/2-\lambda .
\end{equation}
It, therefore, follows from (21), (22) and (47) that

\begin{equation}
\psi _{;ab}u^{a}u^{b}=\alpha (R/3-\rho ),\,\,\psi _{;ab}x^{a}x^{b}=\alpha
(\lambda -R/3),
\end{equation}

\begin{equation}
\psi _{;ab}u^{a}x^{b}=0,\,\,\psi _{;ab}u^{a}h^{bc}=0,\,\,\psi
_{;ab}u^{a}h^{bc}=0.
\end{equation}
Using (48) and $\Box \psi =-\frac{1}{3}\alpha R$ [see Eq.(22)], from
Eqs. (42) and (43) we have

\begin{equation}
\pounds _{_{\xi }}\rho =2\rho (\alpha -\psi ),
\end{equation}

\begin{equation}
\pounds _{_\xi }\lambda =-2\lambda(\alpha+\psi).
\end{equation}
Now, due to (49), Eqs. (44), (45) and (46) reduce to

\begin{equation}
\rho _px^bv_b=0,
\end{equation}

\begin{equation}
\rho h^{ab}v_b=0,
\end{equation}

\begin{equation}
\lambda h^{ab}n_b=0.
\end{equation}
If we consider (52) and (53) and assume that

\begin{equation}
\rho _p\neq 0\,\,\,\,and\,\,\,\,\rho \neq 0
\end{equation}
then from (52) and (53) we have

\begin{equation}
x^bv_b=0,\,\,h^{ab}v_b=0.
\end{equation}
Since $u^bv_b=0\,\,$ we conclude therefore that $v^b\equiv 0$. Equations
(26) and (27) reduce to

\begin{equation}
\pounds _{_\xi }u^a=-\psi u^a,
\end{equation}

\begin{equation}
\pounds _{_\xi }u_a=\psi u_a.
\end{equation}
If we consider (54) and assume that

\begin{equation}
\lambda \neq 0.
\end{equation}
Then we have

\begin{equation}
h^{ab}n_b=0.
\end{equation}
We also have $x^bn_b=0$ and further from (31)

\begin{equation}
u^bn_b=-x^bv_b.
\end{equation}
But if $\rho _p\neq 0$ it follows from (52) that $x^bv_b=0$ and therefore $%
u^bn_b=0$ by (60). Thus since $h^{ab}n_b=0,$ $x^bn_b=0,$ and $u^bn_b$ $=0$
we conclude that$\,\,n_b\equiv 0$. In this case, (28) and (29) reduce to

\begin{equation}
\pounds _{_\xi }x^a=-\psi x^a,
\end{equation}

\begin{equation}
\pounds _{_\xi }x_a=-\psi x_a.
\end{equation}
Also, the derivations of (58) required the reasonable assumptions
contained in (55), and (63) required $\lambda \neq 0$ and $\rho _p\neq 0$%
.

\section{Equations of State}

Let us give an example how the inheritance quantities are useful for
deriving equations of state in the cloud of strings. We will find these for
many special cases, i.e., $\xi $ is parallel to $u^a$ and $x^a$ and
orthogonal to both $x^a$ and $u^a$. We assume a CKV $\xi $ is also a CI. If $%
\xi $ is a CKV satisfying (1), then

\begin{equation}
(R^{ab}\xi _b)_{;a}\,=-3\Box \psi .
\end{equation}
Via Einstein's field equations

\begin{equation}
R^{ab}=T^{ab}-\frac 12Tg^{ab}
\end{equation}
we find

\begin{equation}
[(T^{ab}-\frac 12Tg^{ab})\xi _b];_a=-3\Box \psi .
\end{equation}
Using the Einstein's equations (35), $3\Box \psi =-\alpha R$, equation
(66) reduces to

\begin{equation}
[(T^{ab}-\frac 12Tg^{ab})\xi _b]_{;a}=\alpha R.
\end{equation}
For a cloud of strings with energy-momentum tensor (6) we have

\begin{equation}
(T^{ab}-\frac 12Tg^{ab})\xi _b=\frac 12(\rho -\lambda )u^a(u^b\xi _b)+\frac
12(\rho -\lambda )x^a(x^b\xi _b)+\frac 12(\rho +\lambda )h^{ab}\xi _b.
\end{equation}
First, suppose that $\xi ^a$ is parallel to $x^a$: $\xi ^a=\xi x^a$. Then
since $u^bx_b=0$ and $h^{ab}x_b=0$, (68) reduces to

\begin{equation}
(T^{ab}-\frac 12Tg^{ab})\xi _b=\frac 12(\rho -\lambda )\xi ^a,
\end{equation}
which, when substituted into (67), gives

\begin{equation}
\pounds _{_\xi }\rho -\pounds _{_\xi }\lambda +(\rho -\lambda )\xi _{;a}^a=0.
\end{equation}
But from (1),

\begin{equation}
\xi _{\,;a}^a=4\psi ,
\end{equation}
and using also (50), (51), and $R=\rho +\lambda $, Eq. (50) becomes

\begin{equation}
2\psi \rho =2\psi \lambda .
\end{equation}
From Eq. (72) we have

\begin{equation}
\rho =\lambda .
\end{equation}
Second, suppose that $\xi ^a$ is parallel to $u^a$. Equation (68) reduces to

\begin{equation}
(T^{ab}-\frac 12Tg^{ab})\xi _b=-\frac 12(\rho -\lambda )\xi ^a
\end{equation}
and proceeding as previously we find that

\begin{equation}
\psi (\rho -\lambda )=-2\alpha (\rho +\lambda ).
\end{equation}
Finally, suppose that $\xi ^a$ is orthogonal to $x^a\,$ and $u^a$. Then $%
h^{ab}\xi _b=\xi ^a$ and (68) becomes

\begin{equation}
(T^{ab}-\frac 12Tg^{ab})\xi _b=\frac 12(\rho +\lambda )\xi ^a.
\end{equation}
On substituting (76) into (67) and proceeding as before we obtain

\begin{equation}
\psi (\rho -\lambda )=2\alpha \lambda .
\end{equation}
For $\alpha \neq 0$ (i.e., proper CI), Eq. (77) provide physically
meaningful equation of state for a give $\alpha $ and $\psi $. To illustrate
this point, consider a simple case for which

\begin{equation}
\alpha =\frac w2\psi ,
\end{equation}
where $w$ is constant and $w>0$. Eq. (77) provide the following equation
of state for $\psi \neq 0$

\begin{equation}
\rho =(1+w)\lambda .
\end{equation}
From Eq. (73) we have state equations ($\rho =\lambda $) which coincide
with the state equation for a cloud of geometric (Nambu) strings$^{19}$.
Also, from Eq. (79) we have state equations ($\rho =(1+w)\lambda $) which
coincide with Takabayashi string$^{19}$.

\section{Curvature Inheritance Symmetry
In The Fluids of Strings}

In this section we will consider the fluids of strings described by
energy-momentum tensor (9). With the aid of (27) for $\pounds _{_\xi
}u_a $ and (29) for $\pounds_{_\xi }x_a$ a direct calculation yields

\begin{eqnarray}
\pounds _{_\xi }T_{ab}&=&[\pounds _{_\xi }\rho _s+2\psi \rho
_s]u_au_b-[\pounds _{_\xi }\rho _s+2\psi \rho _s]x_ax_b \nonumber \\
& & +[\pounds _{_\xi }q+2\psi q]h_{ab}+[2\rho _s+2q]u_{(a}v_{b)}-[2\rho
_s+2q]x_{(a}n_{b)},
\end{eqnarray}
which, when substituted into (36), gives

\begin{eqnarray}
2\Box \psi (h_{ab}-u_au_b+x_ax_b)-2\psi _{;ab}&=&[\pounds _{_\xi }\rho_s
+2\psi \rho _s]u_au_b-[\pounds _{_\xi }\rho_s \nonumber \\
& &+2\psi \rho _s]x_ax_b+[\pounds _{_\xi }q+2\psi q]h_{ab} \nonumber \\
& &+[2\rho _s+2q]u_{(a}v_{b)}\nonumber \\
& &-[2\rho_s+2q]x_{(a}n_{b)},
\end{eqnarray}
By contracting (82) in turn with the tensors $u^au^b,\,\,x^ax^b,\,\,h^{ab},%
\,\,u^ax^b,\,\,u^ah^{bc}\,\,$ and $x^ah^{bc}$ following equations are
derived:

\begin{equation}
\pounds _{_\xi }\rho _s+2\psi \rho _s=-2(\Box \psi +\psi _{;ab}u^au^b)
\end{equation}

\begin{equation}
\pounds _{_\xi }\rho _s+2\psi \rho _s=2(\Box \psi -\psi _{;ab}x^ax^b)
\end{equation}

\begin{equation}
\pounds _{_\xi }q+2\psi q=2\Box \psi -\psi _{;ab}h^{ab}
\end{equation}

\begin{equation}
2\psi _{;ab}u^ax^b=0
\end{equation}

\begin{equation}
(\rho _s+q)h^{bc}v_b=\psi _{;ab}u^ah^{bc}
\end{equation}

\begin{equation}
(\rho _s+q)h^{bc}n_b=\psi _{;ab}x^ah^{bc}
\end{equation}
Equations (82)-(87) are valid for any CKV $\xi ^a$ .

From Eq.(85) we have $\psi _{;ab}=0$, i.e., $\xi ^a$ is always a SCKV in
contrast to the cloud of strings. If $\xi ^a$ is a SCKV or HV or KV, then,
there exists no curvature inheritance vector (CIV) other than a curvature
collineation vector (CCV)$^8$. Therefore, string fluid doesn't admit CIV.

\section{Conclusions}

In the case of string cloud, we have found Eqs. (50), (51), (58) and
(63) as inheritance equations. We have shown that the derivations of
(58) and (63) required the reasonable assumptions contained in (55)
and (59), i.e.,$\rho _{p}\neq 0$,$\,\rho \neq 0\,$ and $\lambda \neq 0$.

We have also found under the conditions which $\xi ^a$ is parallel to $x^a$
that there was a relation between $\rho$ and $\lambda$ as the equation of
state ($\rho =\lambda $) for a cloud of geometric strings. Furthermore, we
have found that $\rho =(1+w)\lambda $ which coincide with Takabayashi string
when $\xi ^a$ orthogonal to $x^a$ and $u^a$. For $\alpha=0 $ and $\psi\neq 0
$, we have equation of state ($\rho=\lambda $) from Eqs. (75) and (77).

In the case of string fluids, we have found that $\xi ^a$ is always a SCKV.
So, we conclude that string doesn't admit CIV, i.e., string fluid admits CC
(A CC is also a CKV iff $\psi _{;ab}=0$, see Ref. 7). If $\xi ^a\,\,$is a
SCKV, then we have that

\[
\pounds _{_\xi } q+2\psi q =\pounds _{_\xi }\rho _s+2\psi \rho_s=0,
\]

\[
\pounds _{_\xi }u_a=\psi u_a\,\,\,and\,\,\,\pounds _{_\xi }x_a=\psi
x_a
\]
which is given by Yavuz and Y{\i }lmaz$^6$ as inheritance equations.

\section{Acknowledgment}

The authors would like to thank K. L. Duggal and J. Carot for providing us
with preprints of their papers.

\end{document}